\documentclass{svproc}
\usepackage{xcolor}
\usepackage{amsmath,amssymb}
\usepackage{epsfig}
\usepackage{graphicx,float}
\usepackage{subcaption}
\usepackage{url}
\usepackage[linesnumbered, ruled, vlined]{algorithm2e}
\usepackage{adjustbox}
\usepackage{float}
\usepackage{tikz}
\usepackage{caption} 
\date{}
\begin{document}
\mainmatter

\title{Network analysis and link prediction in competitive women's basketball}

\author{Anthony Bonato\inst{1}\thanks{Supported by an NSERC Discovery Grant.} \and Morganna Hinds\inst{1}}

\institute{Department of Mathematics, Toronto Metropolitan University \\ Toronto, Ontario, Canada}

\maketitle

\begin{abstract}
Network structure and its role in prediction are examined in competitive basketball at the team and player levels. Adversarial game outcome networks from NCAA Division I women’s basketball from 2021 to 2024 are used to compute the common out-neighbor score and PageRank, which are combined into a low-key leader strength that identifies competitors influential through structural similarity despite relatively low centrality. This measure is related to changes in NCAA NET rankings by grouping teams into quantiles and comparing average rank changes across seasons for both previous-to-current and current-to-next transitions. Link prediction is then studied using node2vec embeddings across three interaction settings. 

For NCAA regular-season game networks, cosine similarity between team embeddings is used in a logistic regression model to predict March Madness matchups. For WNBA shot-blocking networks, future directed blocking interactions are predicted via logistic regression on concatenated source-target player embeddings. For WNBA passing networks, region embeddings learned from first-quarter passes are evaluated for their ability to predict subsequent passing connections. Across NCAA and WNBA settings, embedding-based models provide statistically significant evidence that higher-order network structure contains predictive signals for future interactions, while the passing experiment shows weaker predictive performance but yields interpretable similarity patterns consistent with passing feasibility.

\end{abstract}

\section{Introduction}

Sports outcome prediction has long been a central topic in both academic research and commercial analytics, driven by applications in performance evaluation, forecasting, and strategic decision-making. Over time, predictive methodologies have evolved from rule-based and statistical approaches to more data-driven techniques. In recent years, machine learning methods, including neural networks, random forests, and decision trees, have become increasingly prominent, attracting significant attention from both researchers and industry professionals \cite{galekwa2024}.  

Several recent studies have explored the use of machine learning models for predicting basketball game outcomes. In \cite{horvat2020}, seven classification machine learning algorithms were evaluated for NBA outcome prediction, with the $k$-nearest neighbors algorithm achieving the highest accuracy at $60.82\%$. In \cite{zhao2023}, the authors construct a network representation in which teams are modeled as nodes connected to their opponents and to their past and future games. A graph convolutional network trained on this structure was shown to improve predictive performance for NBA game outcomes.

Despite the growing use of machine learning techniques, network centrality measures have received comparatively less attention in sports outcome prediction. Notable exceptions include \cite{xu2018}, which examined Duke University’s men’s basketball passing network by representing players as nodes and passes as edges, and using centrality and betweenness to identify influential players. Similarly, \cite{govan2008} proposes using PageRank on an adversarial National Football League network to effectively rank football teams.

The objective of this paper is to investigate the predictive value of network-based measures and embedding techniques in competitive women's basketball. In particular, we examine the effectiveness of centrality measures and the node2vec algorithmic framework when applied to basketball interaction networks. By modeling competitions, passes, and defensive actions as networks, we aim to show that complex network representations can uncover latent structural information that complements traditional predictive features and offers an alternative perspective on sports outcome prediction. In addition to team-level competition networks, we analyze player-level passing and blocking networks to illustrate how these representations capture interaction patterns that are difficult to observe through aggregate statistics alone.

The paper is organized as follows. Section~2 presents the network-theoretic concepts and embedding-based methods that underpin this work. Section~3 describes the data sources, network construction procedures, and experimental design, followed by the presentation of results. Finally, Section~4 summarizes the main findings and discusses potential directions for future research.

We consider weighted, directed graphs (or \emph{digraphs}) in the paper. Additional background on graph theory may be found in \cite{west}. For more background on complex networks, see \cite{bonato2008}.

\section{Centralities and Embeddings}

To model sports dynamics at both the team and player levels, we represent competitive outcomes and in-game interactions, including passing and blocking, using directed networks that encode relational structure. This network-based perspective enables the application of centrality measures and embedding techniques to quantify influence, similarity, and the underlying structure within the competition.

More formally, an \emph{adversarial network} is a directed graph $G=(V,E)$, where each node represents a competitor, and there is a directed edge $(u,v) \in E$ if $u$ has defeated $v$ in a competition. Edge weights represent the frequency of competitive dominance. 

\subsection{The CON Score}

When predicting competitive outcomes, it is important to identify influential competitors within the network. We adopt a centrality-based measure that quantifies the extent to which a node shares common adversaries with other nodes, which serves as a measure of similarity in competitive behavior and overall influence.

For a given graph $G$, a node $w$ is a \emph{common out-neighbor} of nodes $u$ and $v$ if $(u,w),(v,w) \in E(G)$. We denote by $\mathrm{CON}(u,v)$ the number of common out-neighbors that $u$ and $v$ share. Furthermore, the \emph{CON score}, or \emph{Common Out-neighbor score} \cite{bonato2025}, of a node $u$ is defined as the total number of common out-neighbors that $u$ shares with all nodes in $G$. That is, 
\[\mathrm{CON}(u)=\sum_{v \in V(G)}\mathrm{CON}(u,v).\] For a more in-depth definition of the CON Score, see \cite{bonato2019}.

In basketball competition networks, a node with a high CON score exhibits competitive patterns similar to those of many other nodes. This structural similarity supports informed inference about future performance, as many nodes serve as comparable competitors.

Another commonly used measure of network centrality is \emph{PageRank} (PR). In adversarial networks, PageRank is computed on the reversed-edge network, and nodes with high PageRank values are interpreted as having high centrality. Further details on PageRank and its application to adversarial networks can be found in \cite{bonato2008}.

\subsection{The Low-Key Leader Score}

For a deeper perspective on performance prediction, we are interested not only in influential nodes but also in those that exhibit low centrality. Observe that for any $u$ in a given graph $G$, we have that $\mathrm{CON}(u)$ is a nonnegative integer and $\mathrm{PR}(u) \in [0, 1]$. To compare the two metrics, we apply \emph{unity-based normalization} to both the PageRank and CON scores. This is done by considering the values $X_1, X_2, \dots, X_n$ corresponding to the $n$ nodes in the network and defining \[X_{i,\mathrm{norm}} = \frac{X_i-X_{\mathrm{min}}}{X_{\mathrm{max}}-X_{\mathrm{min}}}.\]
Finally, the \textit{low-key leader strength} \cite{bonato2022} of a node, denoted $\varepsilon_i$, is defined as
\[\varepsilon_i= \mathrm{CON}_{i,\mathrm{norm}}-\mathrm{PR}_{i,\mathrm{norm}}.\] 

A \textit{low-key leader} is a node with a comparatively high low-key leader score. In the context of basketball competition networks, a low-key leader is an ``underdog:'' a competitor who performs similarly to other influential nodes in the network but is less prominent.

\subsection{node2vec} 
Node2vec is an algorithmic framework for learning node feature representations that preserve meaningful neighborhood structure, first introduced in \cite{grover2016}. Let $G = (V, E)$ be a given graph. We define a mapping $f: V \to \mathbb{R}^d$ that assigns each node a $d$-dimensional feature representation, or vector embedding. Equivalently, $f$ may be viewed as a $|V| \times d$ matrix. For each node $u \in V$, $N(u) \subseteq V$ denotes a sampled \textit{network neighborhood} generated according to a sampling strategy $S$. The objective of node2vec is to maximize the likelihood of observing the sampled neighborhood of each node given its embedding. This reduces to a Skip-gram-style formulation, with the context window being the sampled neighborhood, which is then used to optimize $f$ via stochastic gradient ascent.

Node2vec introduces randomized local search procedures to sample multiple neighborhood sets $N(u)$, each of fixed cardinality $k$, for every node $u$. The sampling strategy determines the type of similarity encoded in the embeddings. Two important notions of similarity are:
\begin{enumerate}
\item \emph{Homophily}: Nodes that belong to the same communities or are highly interconnected should be embedded close together.
\item \emph{Structural Equivalence}: Nodes that share similar structural roles (for example, hubs or bridges) should be embedded close together.
\end{enumerate}

In order to accomplish this, we make use of two sampling strategies: 
\begin{enumerate}
    \item \emph{Breadth-first Sampling} (\emph{BFS}): Samples nodes close to current node $u$, yielding a microscopic view of the local structure. This emphasizes \emph{structural equivalence}, since similar roles can be inferred from local neighborhoods.
    \item \emph{Depth-first Sampling} (\emph{DFS}): Samples nodes farther from current node $u$, obtaining a macro-view of the graph. This encourages \emph{homophily} by revealing community structure.
\end{enumerate}

Node2vec interpolates between BFS and DFS through second-order random walks controlled by two parameters, $p$ and $q$. Consider a random walk in which the most recently traversed edge was from node $t$ to node $v$. The probability of moving from $v$ to a neighbor $x$ is proportional to:
\[
\alpha_{p,q}(t,x) \cdot w_{v,x},
\]
where $w_{v,x}$ is the edge weight, and
\[
\alpha_{p,q}(t,x) =
\begin{cases}
  1/p & \text{if } d_{tx} = 0, \\
  1   & \text{if } d_{tx} = 1, \\
  1/q & \text{if } d_{tx} = 2.
\end{cases}
\]
where $d_{tx}$ is the shortest path distance between $t$ and $x$.

Thus, $p$ acts as a \emph{return parameter}, controlling the likelihood that a random walk revisits the previously visited node. Larger values of $p$ reduce the probability of immediately returning to the previous node, while smaller values of $p$ increase the tendency to backtrack and remain close to the starting node. Parameter $q$ acts as an \emph{in-out parameter}, controlling the likelihood that a random walk explores outward from the previous node versus remaining within its local neighborhood. Larger values of $q$ bias the walk toward breadth-first exploration, while smaller values of $q$ encourage depth-first exploration. 

By tuning $p$ and $q$, node2vec balances local and global exploration, allowing embeddings to capture both structural equivalence and homophily. Random walks are computationally efficient and enable sample reuse, allowing neighborhoods to be generated for many nodes in a single walk. In this paper, we implemented node2vec with $p=1$ and $q=1$, the default values of $p$ and $q$.

\section{Experimental Design and Methods}

\subsection{NCAA Women's Basketball Game Data}
An average of 355 American college women's basketball teams participated in the NCAA Division 1 Women's basketball league from 2021 to 2024. The teams, each belonging to 1 of the 31 conferences, play non-conference games early in the regular season, and conclude the regular season with in-conference games and a tournament.

The \emph{NCAA March Madness tournament} is a single-elimination competition held after the regular season concludes. A total of 68 teams participate and compete for the national title. The 31 winners of the conference tournaments receive an automatic bid to March Madness, meaning they have a reserved spot in the tournament. The NCAA basketball committee votes on its selections for the other 37 tournament positions. The basketball committee then undertakes an extensive process to create a tournament schedule that aims to achieve relative balance. This entails splitting the teams into four regions based on the committee's rankings, while attempting to separate same-conference teams as much as possible. Figure~1 demonstrates the structure of each region's bracket. The winners of the \emph{Elite Eight} games enter the \emph{Final Four}, where the region 1 winner faces the region 4 winner, and the region 2 winner faces the region 3 winner. Finally, the winners of the Final Four games compete in the \emph{National Championship} game. For further information on the March Madness procedures, refer to \cite{ncaa2025principles}. 
 
\begin{figure}[htpb!]
  \centering
  \includegraphics[
    width=0.7\textwidth,
    height=0.315\textheight,
    keepaspectratio,
  ]{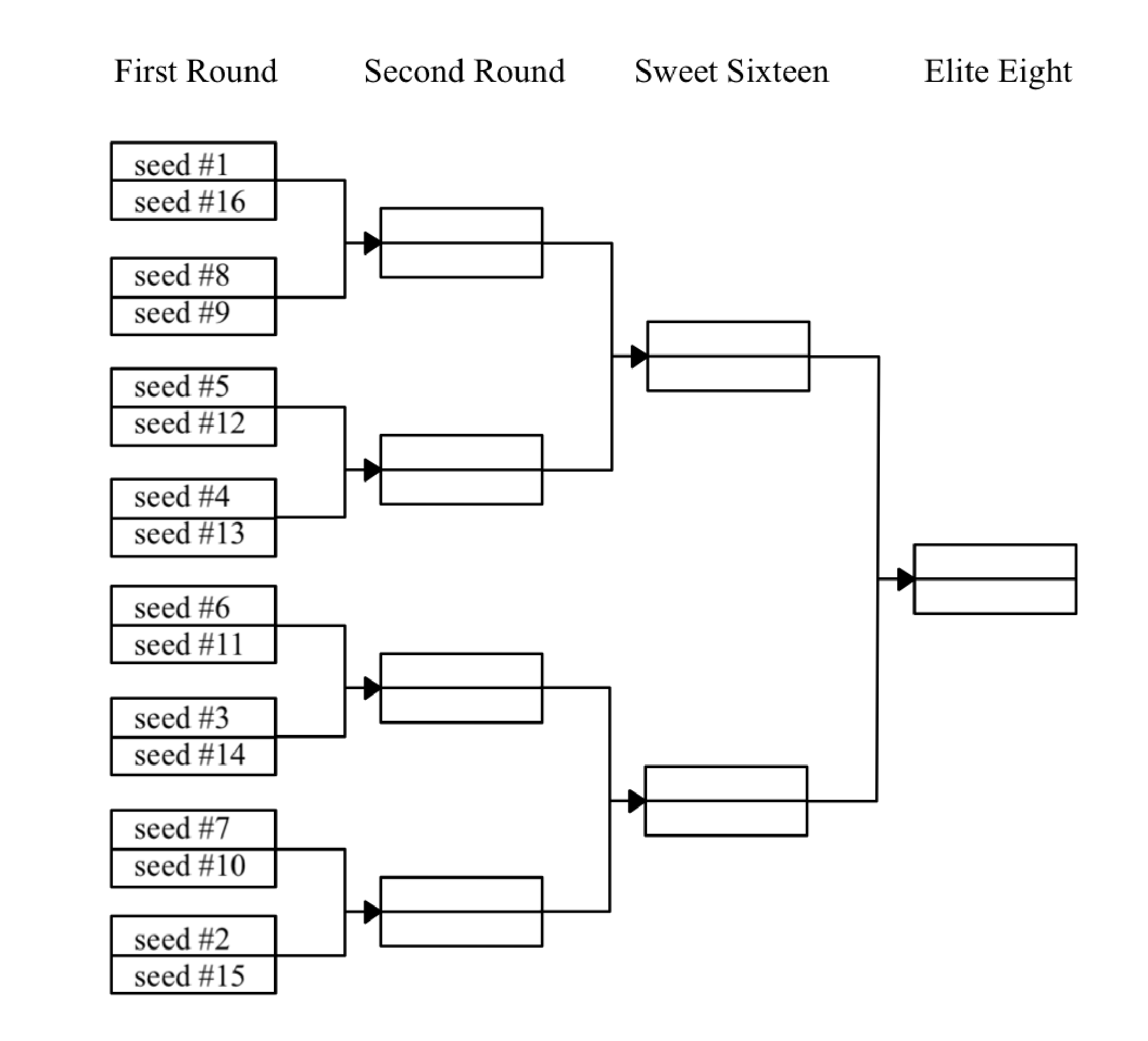}
  \caption{The structure of a March Madness regional bracket.}
  \label{fig:mm-bracket}
\end{figure}

Using a comprehensive dataset from Kaggle \cite{kaggle2025mania}, we acquired the final scores of every basketball game played within NCAA Division 1 Women's basketball from 2021-2024. Each season averaged 4917 total games. A game data network $G$ was constructed for each season, with nodes representing teams. Scoring data between each pair of teams was totaled throughout their matchups for the entire season, and $(u,v) \in E(G)$ if team $u$ outscored team $v$ across all their head-to-head games in the season, with the total amount outscored represented by edge-weight. Figure~2 shows the network formed from both Regular season and March Madness game data for the 2023-2024 season, consisting of 360 nodes and 3903 edges.

\begin{figure}[htpb!]
  \centering
  \includegraphics[
  width=0.7\textwidth,
  height=0.315\textheight,
  keepaspectratio
]{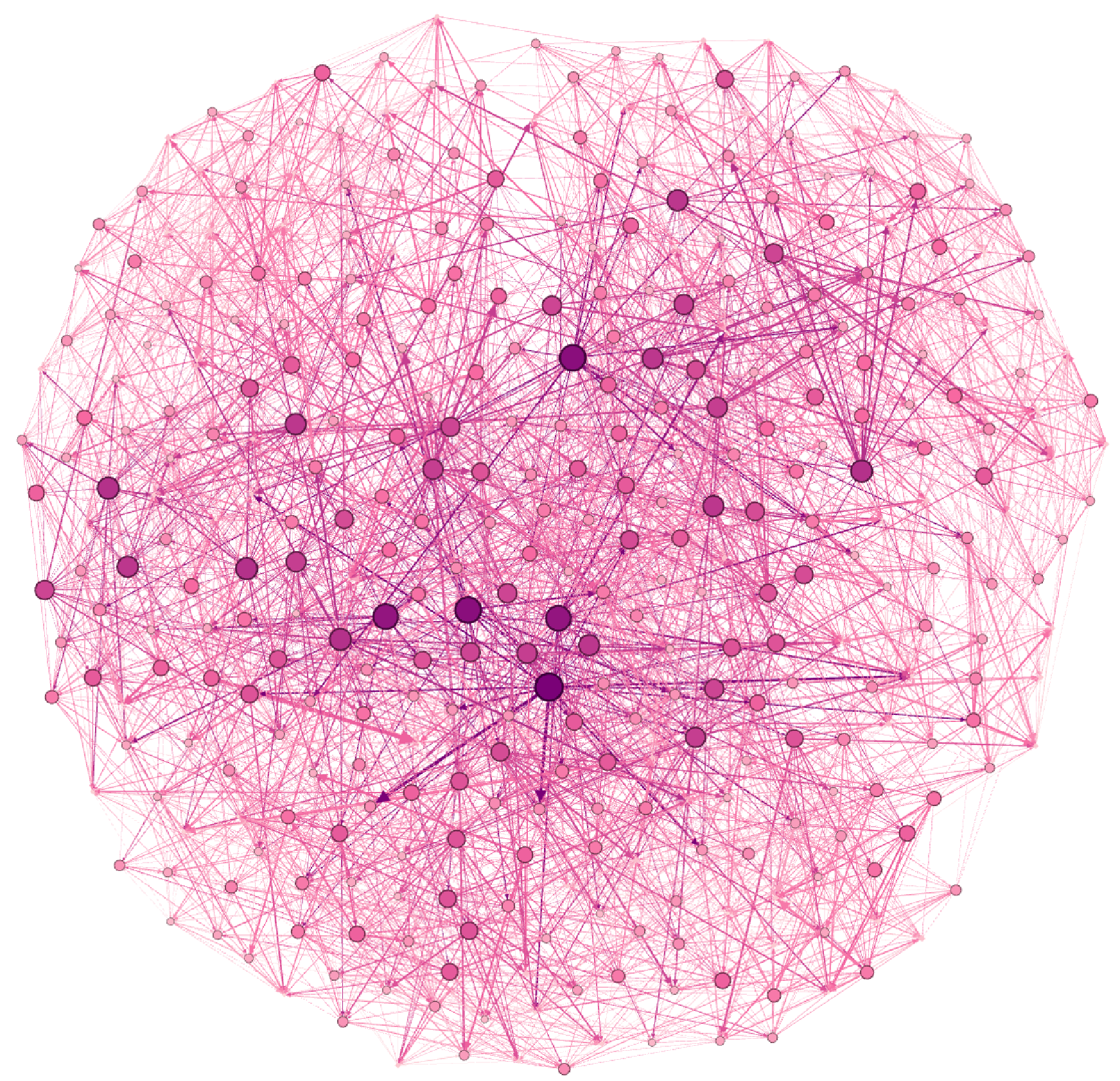}
  \caption{An NCAA 2023-2024 season network. Nodes with higher comparative out-degree are larger and darker.}
  \label{fig:mm-bracket}
\end{figure}

\subsection{WNBA Passing Data}

The \emph{Women's National Basketball Association (WNBA)} is a professional basketball league featuring the world's top female players. Each game consists of four 10-minute quarters, with an additional overtime segment if the score is tied after all quarters. During games, players dribble and pass the basketball to teammates to create open shots. Ball movement is essential to create scoring opportunities for a team. Accordingly, it is imperative that teams agree on pass sequences to optimize scoring. These sequences of passes are called \emph{plays}, and different plays are strategically used throughout the game. 

We viewed and manually coded passing data from publicly available WNBA game broadcasts for three WNBA games: Valkyries versus Lynx (09/14/25), Mercury versus Liberty (09/14/15), and Aces versus Fever (09/21/25). Splitting the court into 28 distinct regions, we documented the locations of every pass, both source and target, during these games. The result was an average of 365 passes per game. For each game, we used its corresponding passing data to construct a directed network $G$, where the nodes are regions of the court, and $(u,v) \in E(G)$ if a pass from region $u$ to region $v$ occurred during the game with $weight(u,v)$ being the number of occurrences. To our knowledge, this is the first time that sports passing data has been represented in this way.

Figure~3 represents the network formed by the 317 total passes made during the Aces vs. Fever game. Note that front-court passes for both teams are tracked on the left-hand side of the network, and back-court passes on the right. 

\begin{figure}[H]
  \centering
  \includegraphics[
  width=\textwidth,
  height=0.5\textheight,
  keepaspectratio
]{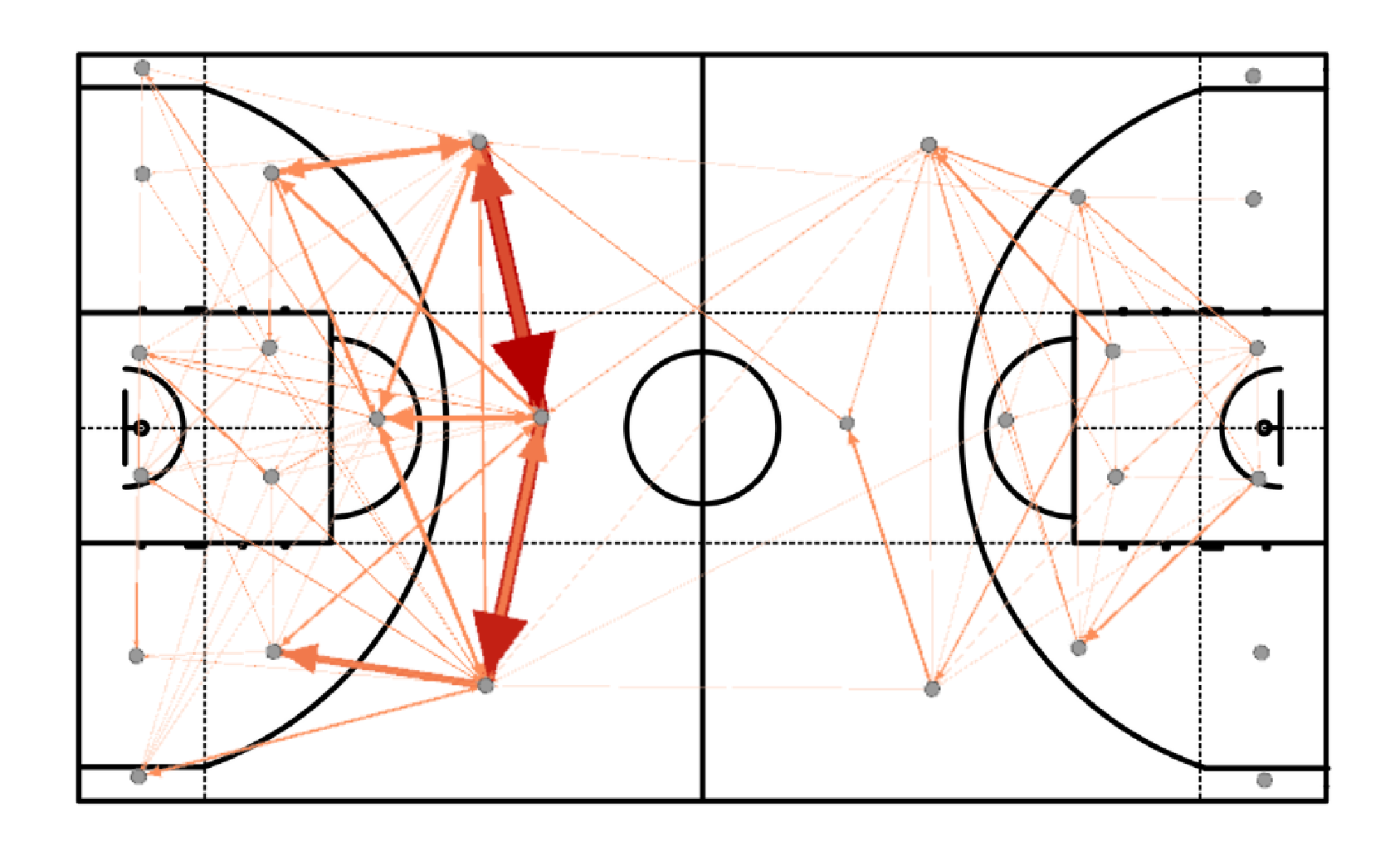}
  \caption{Aces vs.\ Fever 09/21/25 passing network. Edges of higher weight are larger and darker in colour.}
  \label{fig:mm-bracket}
\end{figure}

\subsection{WNBA Blocking Data}

A crucial defensive tactic in the WNBA is shot-blocking. \emph{Shot-blocking} (or more simply, \emph{blocking}) \cite{wnba2022rules} occurs when a defensive player prevents another player's shot using their arms. This is legal so long as they do not move into the shooter's line of fire. Typically, most blocks are made by players who play the forward position, as they are taller. The source and target of the block depend on many variables, including the court's location and the plays in progress. 

Using shot-blocking data from PBP Stats \cite{pbpstats}, we collected the source and target of every shot block that occurred in the 2023 and 2024 WNBA seasons. There was an average of 1923 blocks per season, with the top-blocking player, A'ja Wilson, having blocked 75 shots in a season. We used this data to form networks $G_1$ and $G_2$ for the 2023 and 2024 seasons, respectively. The nodes in each network are players, and $(u,v) \in E(G)$ if player $u$ blocked player $v$ more than vice versa, with $weight(u,v)$ measuring the number of blocks. Figure 4 shows $G_2$, the network constructed using data from the 2024 season. 

\begin{figure}[H]
  \centering
  \includegraphics[
  width=\textwidth,
  height=0.5\textheight,
  keepaspectratio
]{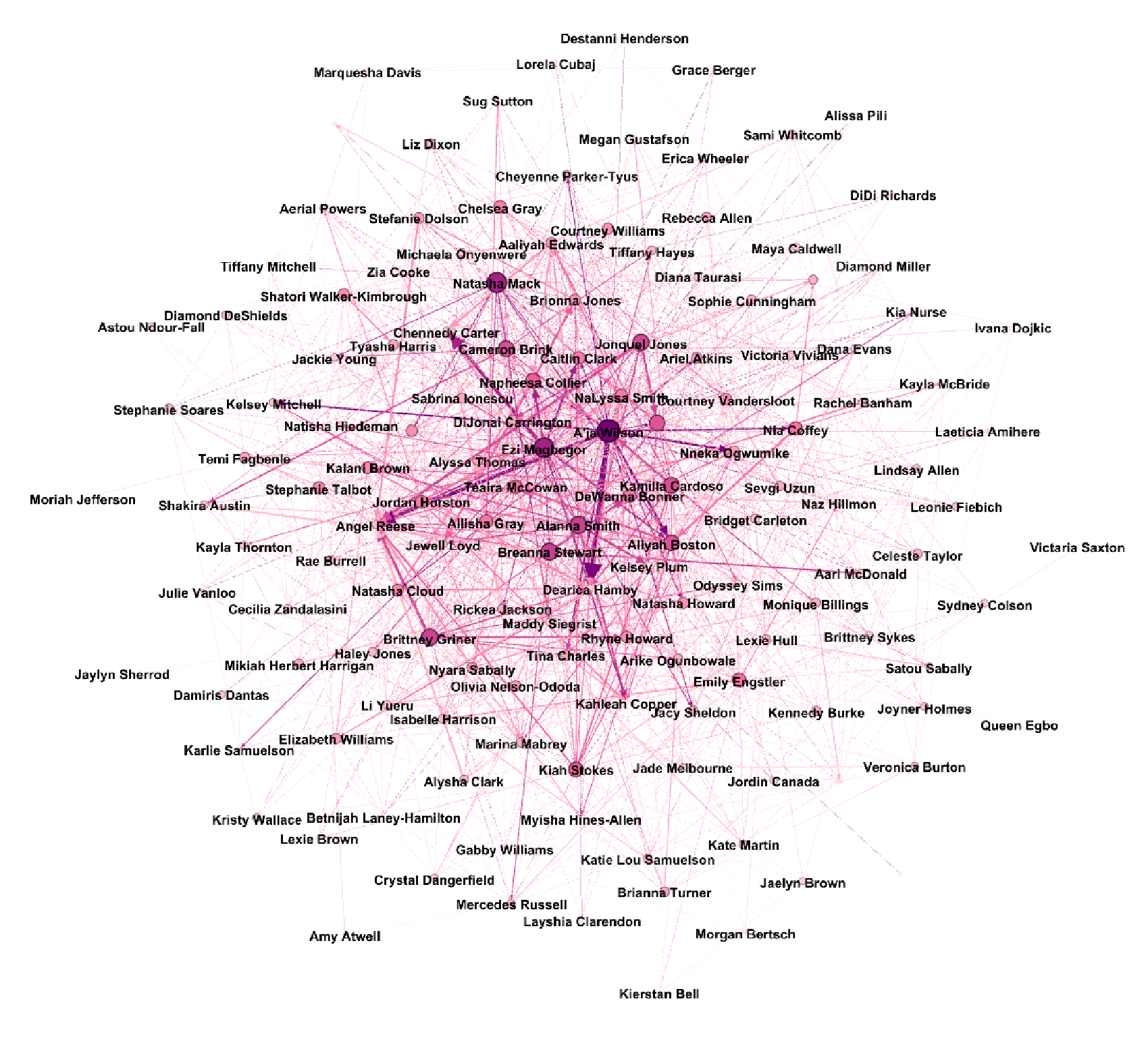}
  \caption{WNBA 2024 season blocking network. Nodes with higher comparative out-degree are larger and darker.}
  \label{fig:mm-bracket}
\end{figure}

\subsection{Results}
Using NCAA women’s basketball competition networks from the 2021–2024 seasons, we computed the CON score, PageRank, and the resulting low-key leader strength for each node. All computations were performed using custom Python scripts developed for this analysis, which are available at \url{github.com/morgannahinds}. The resulting low-key leader strengths for NCAA Division I women’s basketball teams ranged from $-0.361$ to $0.949$ across the seasons considered.

The low-key leader scores were then compared with each team’s \emph{NCAA Evaluation Tool (NET) ranking}, the NCAA’s primary team evaluation metric. Ranking data were scraped from \url{warrennolan.com} \cite{warrennolan}. For additional details on the methodology used to compute NET rankings, see \cite{ncaaNET2025}. We hypothesized that teams with higher relative low-key leader strength would show improvements in NET ranking from the previous to the current season, whereas teams with lower low-key leader strength would experience declines in NET ranking over the same period. Teams were grouped into quantiles based on low-key leader strength, and the average change in ranking from the previous season to the current season was computed for each quantile. The results for the 2024 low-key leader strengths, shown in Figure~4, are consistent with this hypothesis.

\begin{figure}[H]
  \centering
  \includegraphics[
  width=\textwidth,
  height=0.5\textheight,
  keepaspectratio
]{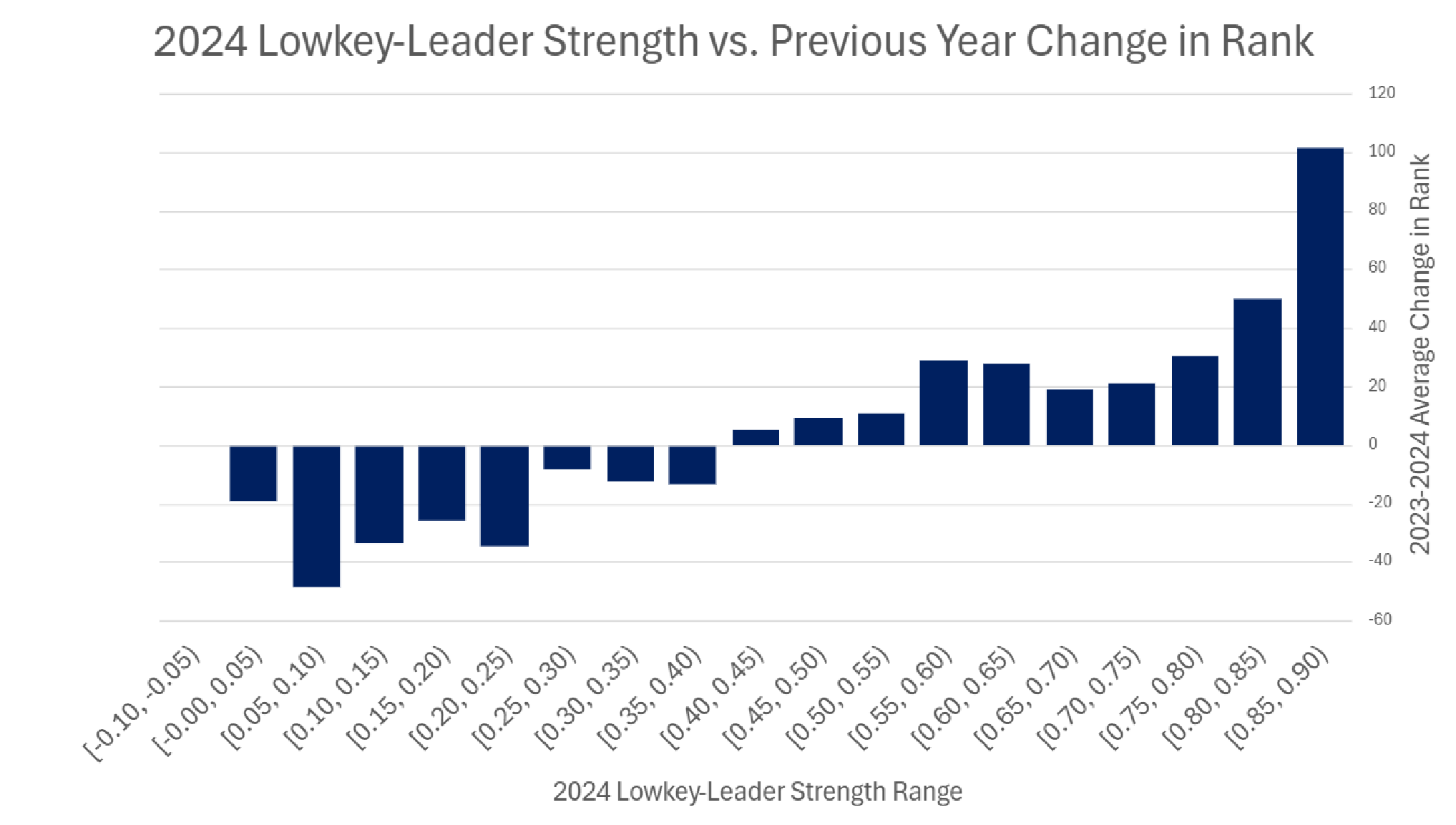}
  \caption{Average change in NET Ranking from 2023-2024, based on 2024 low-key leader strength.}
  \label{fig:mm-bracket}
\end{figure}

We compiled corresponding results for the 2022, 2023, and 2024 seasons. Defining low-key leaders as teams with low-key leader strength in the interval $[0.4,1]$, we found that 58 of the 61 quantiles satisfied the stated hypothesis. The same methodology was applied to NCAA men’s basketball networks over the same seasons, using game data from \cite{kaggle2025mania} and NET rankings from \cite{warrennolan}. In this setting, 50 of the 59 quantiles satisfied the hypothesis. The full dataset, including low-key leader strengths and average changes in ranking, is available at \url{github.com/morgannahinds}.

To assess the predictive potential of low-key leader strength for future rankings, we extended the analysis by evaluating changes in NET ranking from the current season to the following season. We hypothesized that higher low-key leader strength would be associated with declines in future NET ranking, whereas lower low-key leader strength would be associated with improvements. The results based on the 2022 low-key leader strengths are shown in Figure~5.

\begin{figure}[H]
  \centering
  \includegraphics[
  width=\textwidth,
  height=0.5\textheight,
  keepaspectratio
]{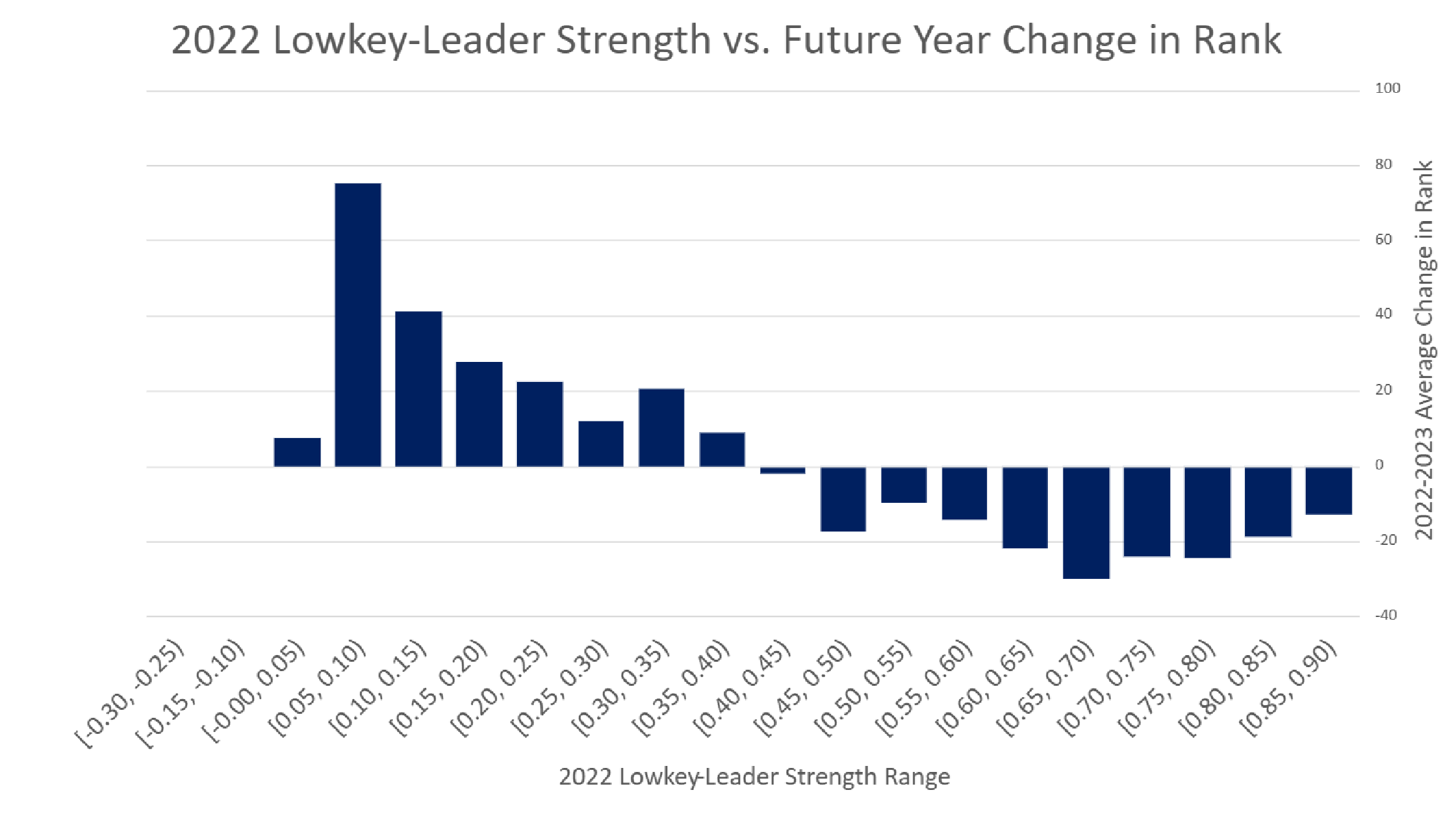}
  \caption{Average change in NET Ranking from 2022-2023, based on 2022 low-key leader strength.}
  \label{fig:mm-bracket}
\end{figure}

While the low-key leader analysis focuses on ranking-level prediction, we also investigated whether latent network structure can be leveraged to directly predict future interactions. To this end, we applied the node2vec algorithm to each network to learn low-dimensional node representations for link prediction. Network construction and embedding generation were implemented in Python using NetworkX, with node2vec applied via the node2vec Python package \cite{node2vecpypi}. Hypotheses were evaluated by fitting logistic regression models using the statsmodels module, and statistical significance was assessed across 100 independent iterations to account for the stochastic nature of the random walks underlying node2vec. 

We first applied node2vec to the NCAA game data networks for the 2022, 2023, and 2024 seasons, considering only regular-season games. We hypothesized that teams with similar embeddings would be more likely to face each other in the subsequent March Madness tournament. Embedding similarity was quantified using cosine similarity, and a logistic regression model was constructed using this similarity measure as a predictor of observed March Madness matchups. Table~1 reports the average model statistics across all 100 iterations for each season.

\begin{table}[H]
    \centering
    \renewcommand{\arraystretch}{1.2} 
    \begin{tabular}{|l|c|c|c|}
        \hline
        \textbf{Statistic} & \textbf{2022 Value} & \textbf{2023 Value} & \textbf{2024 Value} \\ \hline
        \# team pairs        & 2152               & 2135                & 2142            \\ 
        \# embedding features        & 1             & 1             & 1         \\ 
        Pseudo $R^2$       & $1.87 \times 10^{-2}$                & $1.2 \times 10^{-2}$                 & $3.16 \times 10^{-2}$           \\ 
        LLR p-value & $2.36 \times 10^{-3}$                & $1.69 \times 10^{-2}$                 & $4.16 \times 10^{-5}$              \\ 
        Significant embedding dimensions ($p < 0.05$)        & 1       & 1          & 1      \\ 
         \hline
        \end{tabular}
    \caption{Logistic regression results using cosine similarity metrics of node2vec embeddings from the regular season game data network to predict March Madness interactions.}
\end{table}

Logistic regression models using cosine similarity of node2vec embeddings from regular-season NCAA game networks provide statistically significant evidence that cosine similarity is an effective predictor of March Madness matchups across the 2022–2024 seasons. In each year, the likelihood ratio test rejects the null model ($p < 0.05$), indicating that embedding similarity improves model fit. Although the pseudo-$R^2$ values are modest, this is expected given the sparsity and unpredictability of tournament pairings. Overall, these results suggest that teams that are close together in the regular season are more likely to meet in March Madness.

We next applied the node2vec algorithm to the 2023 WNBA blocking network to determine whether player embeddings could predict future blocking interactions in the 2024 season. Using blocking data from the 2023 WNBA season, we generated 10-dimensional player embeddings and tested their predictive utility by fitting a logistic regression model with the embeddings as predictors of binary blocking outcomes in 2024. Because similar embeddings indicate players with comparable attributes, and blocking interactions are inherently asymmetric, cosine similarity was not appropriate in this setting. Instead, we constructed predictor vectors by concatenating the source and target players' embeddings. Table~2 reports the average summary statistics across all iterations.

\begin{table}[H]
\centering

\label{tab:n2v_logit_summary}
\begin{tabular}{|l|c|}
\hline
\textbf{Statistic} & \textbf{Value} \\
\hline
\# player pairs & 5356 \\
\# embedding features & 20 \\
Pseudo $R^2$ & 0.022 \\
LLR p-value & $4.17 \times 10^{-3}$\\
Significant embedding dimensions ($p < 0.05$) & 11 of 20 \\
\hline
\end{tabular}
\caption{Logistic regression results using concatenated node2vec embeddings
from the 2023 blocking network to predict 2024 blocking interactions.}
\end{table}

The logistic regression model using concatenated node2vec embeddings from the 2023 blocking network provides statistically significant evidence that the underlying network structure predicts blocking interactions in the 2024 season. The likelihood ratio test strongly rejects the null model ($p = 0.004$), indicating that the embedding-based predictors jointly improve model fit. Again, the pseudo-$R^2$ value is modest (0.022), likely due to the sparsity and stochastic nature of blocking events. Several embedding dimensions are individually significant, with both positive and negative coefficients, suggesting that multiple structural factors contribute to future blocking behavior. Overall, these findings support the conclusion that higher-order network structure, as captured by node2vec embeddings, contains meaningful information about the formation of future blocking interactions.

We also applied the node2vec algorithm to the WNBA passing network $G$ to assess whether region-level embeddings can predict future passing behavior. In this setting, embeddings were learned from a subgraph of $G$ containing only first-quarter passing data, and we hypothesized that pairs of regions with high cosine similarity would be more likely to be connected by an edge in subsequent quarters. While this approach did not yield strong predictive performance for future pass formations, the resulting similarity scores provided meaningful qualitative insight into the feasibility of passing. Figure~7 illustrates the cosine similarity results of the Fever (09/21/25) for two representative nodes: node 3, the most prominent passing region in the network, and node 13, the least prominent.

\begin{figure}[h!]
    \centering
    \begin{minipage}{0.45\textwidth}
        \centering
        \includegraphics[width=\textwidth]{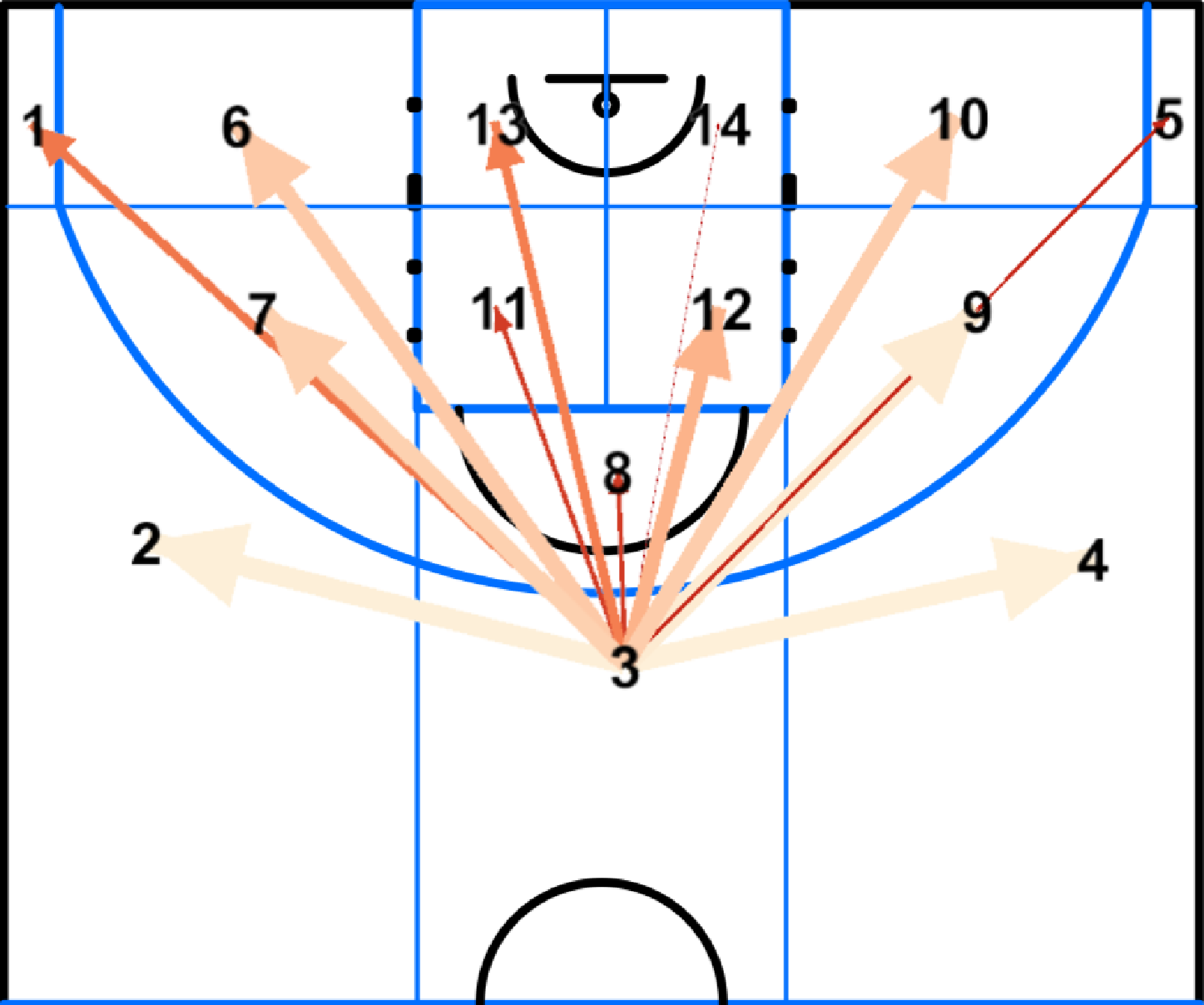}
        \caption*{(a) Node 3}
    \end{minipage}
    \hfill
    \begin{minipage}{0.45\textwidth}
        \centering
        \includegraphics[width=\textwidth]{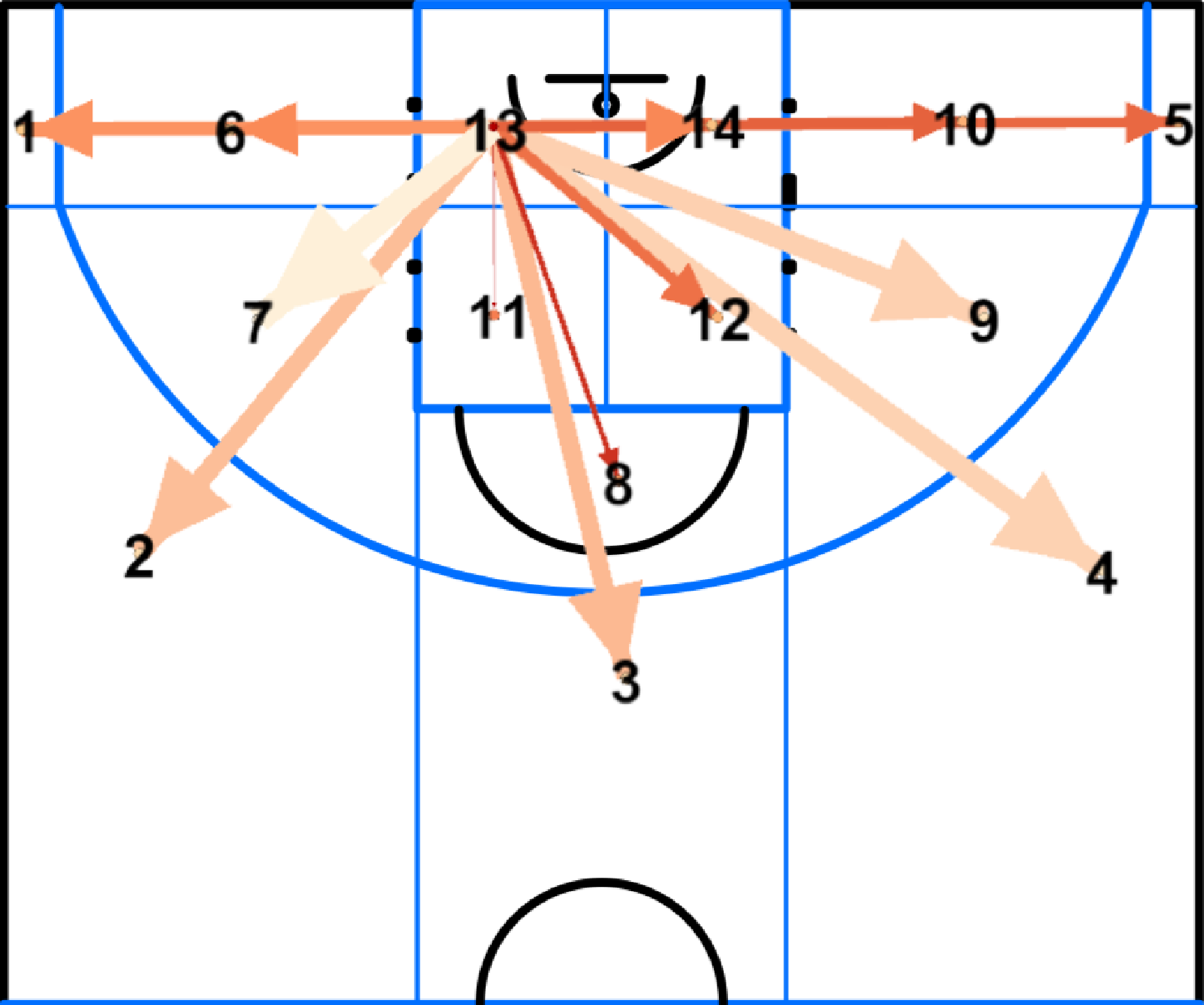}
        \caption*{(b) Node 13}
    \end{minipage}
    \caption{Fever 09/21/25 passing likelihood from nodes 3 and 13 based on cosine similarity of region embeddings. Edges with higher likelihood are larger and lighter in colour. In the Fever passing network, node 3 had the largest out-degree (27) and node 13 had the least out-degree (0).}
\end{figure}

Figure~7a provides a clear categorization of realistic passes originating from region 3. Regions 1 and 5 represent unrealistic passing options due to distance. Although proximity generally increases pass feasibility, several nearby regions present exceptions. Regions 8, 11, 12, 13, and 14 are classified as \emph{high-traffic areas}, meaning that they typically contain multiple players from both teams. This congestion substantially increases the difficulty of passing from region 3 into these areas. In contrast, passes to regions 2, 4, 6, 7, 9, and 10 are more feasible, as these regions are \emph{low-traffic areas} with fewer defenders present. Among these, regions 2, 4, and 9 are particularly favorable due to their proximity and the predominance of right-handed passing mechanics.

Figure~7b illustrates a more challenging passing scenario originating from region 13. Owing to the confined nature of this region and its location within a high-traffic area, passes to nearby high-traffic regions, specifically regions 8, 11, 12, and 14, are likely to be difficult. Passes to regions 1, 5, 6, 10, and 14 are largely infeasible due to the proximity of the out-of-bounds line and the ability of defenders to control the remaining space. In contrast, passes to regions 2, 3, 4, 7, and 9 are more viable options, as these low-traffic regions are sufficiently distant to loft the ball over defenders.

\section{Discussion and Future Directions}

In this work, we demonstrated how network-based representations of competitive and cooperative interactions in basketball can be leveraged to inform performance analysis and link prediction across both collegiate and professional contexts. Using adversarial game outcome networks, we showed that low-key leader strength captures meaningful structural information about changes in NCAA NET rankings. In parallel, we applied node2vec embeddings to multiple basketball networks and found statistically significant evidence that latent network structure predicts future interactions, including postseason NCAA matchups and WNBA shot-blocking behavior, despite the inherent sparsity and stochasticity of these events. While embedding-based prediction was less effective for temporal pass formation in WNBA games, similarity measures still provided interpretable insight into spatial decision-making and passing feasibility. Collectively, these results highlight the value of combining centrality-based measures with embedding-based machine learning approaches to capture higher-order structural signals in sports networks.

From a sports analytics and applied machine learning perspective, these findings indicate that network-derived features, particularly centrality-based measures, can capture underlying competitive information that is not fully reflected in traditional ranking systems. At present, such features remain underutilized in mainstream sports prediction pipelines, which typically emphasize box-score statistics, efficiency metrics, or traditional rating-based models. The low-key leader score provides a compact, interpretable network feature that can be incorporated alongside standard covariates in team-level predictive models, rather than relying solely on quantile-based analysis as introduced in this paper. 

The relationship between low-key leader score and changes in ranking was also evaluated using college football game outcome networks for the 2022–2024 seasons. Game data were obtained from the CollegeFootballData API \cite{cfbdpython}, with rankings collected from \cite{warrennolan}. Defining low-key leaders as teams with low-key leader strength in the interval $[0.4,1]$, we found that 50 of 58 quantiles satisfied the same directional hypothesis observed in NCAA men’s and women’s basketball. Notably, no violations occurred within the interval $[0.6,1]$, suggesting that stronger low-key leader signals correspond to more stable predictive behavior. 

In contrast, comparable performance was not observed in structurally dissimilar sports such as college baseball and professional tennis. Basketball and football are both physical, team-based sports in which players have specialized roles yet remain free to move throughout the entire playing surface, unlike baseball or tennis, where movement is either highly constrained or individual. This combination of structural freedom and continuous interaction may help explain why centrality measures behave similarly in basketball and football, while proving less effective in college baseball and professional tennis. For the same reasons, these techniques are likely to extend to other structurally similar sports such as soccer, rugby, and hockey.

Model performance may be further improved by integrating contextual variables such as home–away effects, injuries, roster continuity, or schedule strength. In addition, increased access to fine-grained spatiotemporal data, potentially enabled by computer vision, would enable the application of network embeddings and centrality measures to passing- and play-structure networks at scale. In this setting, machine learning models could jointly leverage offensive and defensive network representations to support matchup-specific prediction, simulation, and strategic analysis.

\end{document}